\documentclass[prd,nofootinbib,showpacs,preprint]{revtex4}
\usepackage{graphicx}
\usepackage{epsfig}
\usepackage{amsmath}
\usepackage{amsfonts}
\usepackage{amssymb}
\usepackage{url}
\usepackage{hyperref}
\usepackage{subfigure}
\usepackage{cancel}
\usepackage{eso-pic}
\usepackage[abs]{overpic}
\usepackage{xcolor,varwidth}
\newcommand{\beqa}{\begin{eqnarray}}
\newcommand{\eeqa}{\end{eqnarray}}
\newcommand{\beq}{\begin{equation}}
\newcommand{\eeq}{\end{equation}}

\newcommand{\bmt}{\begin{pmatrix}}
\newcommand{\emt}{\end{pmatrix}}
\usepackage[toc,page]{appendix}
\usepackage{comment}
\newcommand{\be}{\begin{equation}}
\newcommand{\ee}{\end{equation}}
\newcommand{\bea}{\begin{eqnarray}}
\newcommand{\eea}{\end{eqnarray}}

\usepackage{xspace}

\begin{document}
\title{Tracking Detector Performance and Data Quality in the NOvA Experiment}
\author{Biswaranjan Behera}
\email{bbehera@fnal.gov}
\affiliation{\,Department of Physics, IIT Hyderabad,
              Telangana - 502285, India \\
Fermilab, USA\\On behalf of the NOvA Collaboration\\ \\
              Talk presented at the APS Division of Particles and Fields Meeting (DPF 2017), July 31-August 4, 2017, Fermilab. C170731}

\begin{abstract}
NOvA is a long-baseline neutrino oscillation experiment. It uses the NuMI beam from Fermilab and two sampling calorimeter detectors located off-axis from the beam. The NOvA experiment measures the rate of electron-neutrino appearance in the almost pure muon-neutrino NuMI beam, with the data measured at the Near Detector being used to accurately determine the expected rate at the Far Detector. It is very important to have automated and accurate monitoring of the data recorded by the detectors so any hardware, DAQ or beam issues arising in the 344k (20k) channels of the Far (Near) detector which could affect the quality of the data taking are determined. This paper will cover the techniques and detector monitoring systems in various stages of data taking.
\end{abstract}
\maketitle
\section{Introduction}
\subsection{NOvA Experiment}
\begin{figure}[!htbp]
    \centering
   \includegraphics[width=0.67\textwidth]{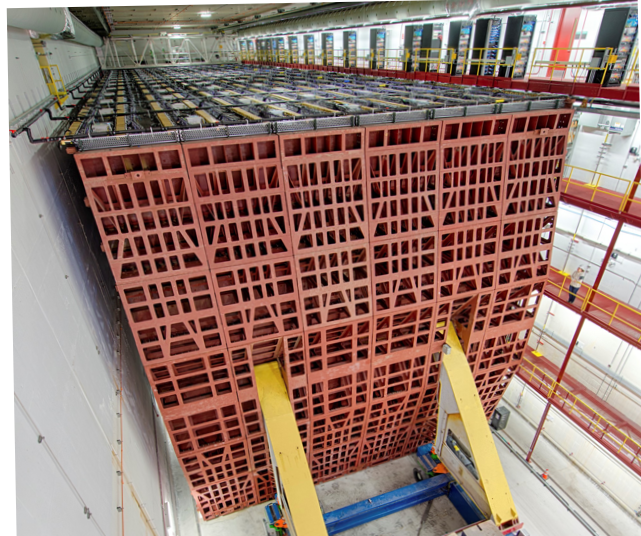}
    \includegraphics[width=0.3\textwidth]{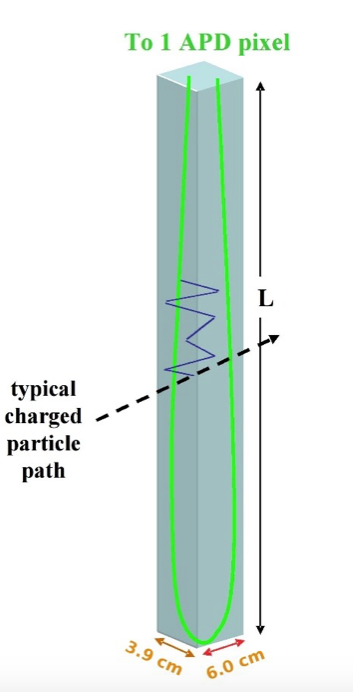}
    \caption{Diagram of the NOvA far detector (left). Schematic diagram of NOvA cell, the walls are made up of PVC with a loop of wave-length shifting fiber is read out by one APD (right).}
   \label{fig:fardet}
  \end{figure}

\begin{figure}[!htbp]
    \centering
   \includegraphics[width=0.6\textwidth]{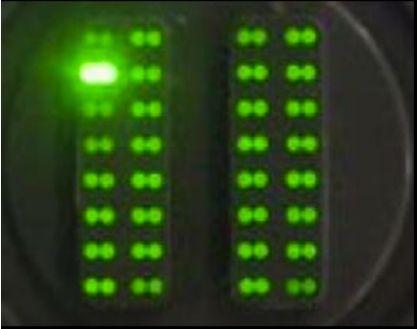}
    \includegraphics[width=0.39\textwidth]{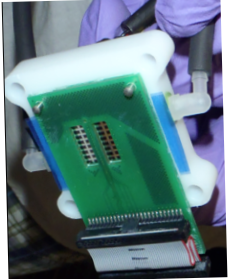}
    \caption{The ends of 32 wavelength-shifting fibers collected at the end of scintillation cells to mount to an APD (left).  The front face of an APD will be pressed against the fiber ends (right).}
   \label{fig:apd}
  \end{figure}

NOvA (NuMI Off-axis $\nu_{e}$  Appearance) is the US flagship long-baseline neutrino oscillation experiment \cite{tdr}. The muon neutrinos beam from the NuMI accelerator at Fermilab passes through two functionally identical tracking detectors, one located at Fermilab, the other 810 km away at Ash River, MN. The two detectors are  placed 14.6 mrad off-axis from the center of the neutrino beam and their similarity in their physical structure and architecture of the data acquisition help in reducing the impact of systematic uncertainties. Both detectors are made of with a reflective PVC tubes (cells) (Fig.\ref{fig:fardet}, right) that is filled with liquid scintillator. Each cell contains one fiber, corresponding to one pixel of the detector and 32 cells make up one module. A series of modules glued together form a plane (1 plane = 12 modules in Far Detector (Fig.\ref{fig:fardet}, left) and 3 modules in Near Detector (Fig.\ref{fig:neardet}, left)). The cells are 3.6 cm by 5.6 cm in cross section and 4.2 m in length in the near detector (ND) and 15.2 m long in the far detector (FD). Cells are stacked in alternating horizontal and vertical planes to provide three dimensional tracking of particles. Wavelength-shifting fiber is looped through inside the cell and collects light from the interaction of a charged particle with the liquid scintillator. The fiber ends terminate on a single pixel of an avalanche photodiode (APD) (Fig.\ref{fig:apd}). The APDs are a critical component of the experiment, as they convert the photons produced by charged particles traversing the scintillator in the PVC tubes of the detectors into electronic signals. The signal is amplified and read out by a Front End Board (FEB) of which there is one for each APD. The FEB digitizes the hits above threshold and reads the analog signal from the APD. The hit information from the FEB is collected by a Data Concentrator Module (DCM). DCMs passes the data to a processing farm. 

\subsection{Far Detector}
 In total there are 896 planes with 344,064 channels and a total active detector mass of 14 kt (65\% liquid scintillator). The detector was constructed in 14 di-blocks (two set of 32 plane blocks glued together). Each di-block has 12 DCMs, with 64 FEBs in each DCM.  
The FD is on the surface and receives 50-70 cosmic rays in each 550 microsecond readout window. Its primary goal to measure the energy spectra for our beam neutrinos, separating muon and electron neutrino charged-current interactions from neutral-current interactions. 

\subsection{Near Detector}
\begin{figure}[!htbp]
    \centering
\includegraphics[width=0.67\textwidth]{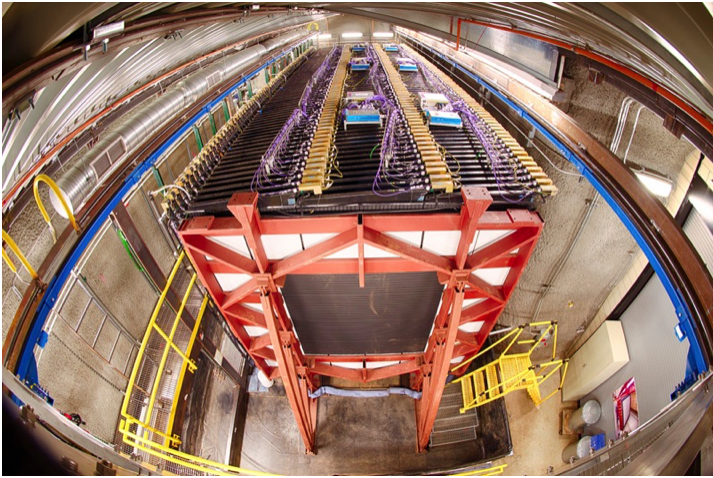}
    \includegraphics[width=0.3\textwidth]{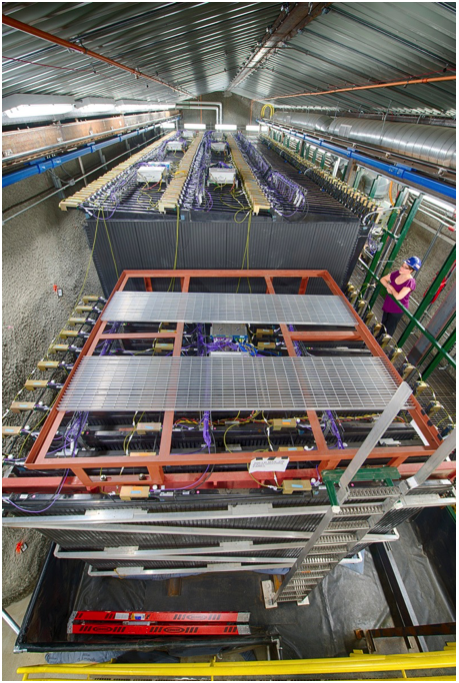}
    \caption{NOvA near detector (left) and muon catcher region (right) with steel plates alternating with scintillation planes, whose height is two-thirds that of the active region.}
 
   \label{fig:neardet}
  \end{figure}

The ND is $\sim$100 m underground to reduce cosmic ray interactions. It is only 1 km away from the beam, so the neutrino interaction rate is much higher than the far detector. It is designed to measure the unoscillated beam spectra so that it can be used to predict the far detector spectrum. The dimension of detector is smaller in size compared to FD. The length of detector is 15.9 meters long along the beam direction, divided into a 12.67 meter active region followed by a 3.23 meter muon catcher at the downstream end as shown in Fig.\ref{fig:neardet} (right). The muon catcher consists of alternating steel planes and scintillating planes to increases the efficiency to contain muons. The ND totals 290 tons, 130 tons of which is liquid scintillator. The active region has mass of 193 tons  with $\sim$18k channels and 192 planes followed by 22 planes and 10 steel planes which are 10 cm thick. The dimension of a detector in the active region is 3.9 $\times$ 3.9 $\times$ 12.67 (h $\times$ w $\times$ l) meters. The active region electronics are instrumented in 3 di-blocks. Each di-block has two DCMs for the vertical planes and two for the horizontal planes. One DCM is each view is fully occupied with 64 FEBs and the other is half occupied with 32. The muon catcher region is a repeated sequence of horizontal plane, steel, and vertical planes. There is one DCM for the vertical modules and one for the horizontal in the muon catcher.
  
  
\section{Performance of the NOvA Detectors}
\begin{figure}[!htbp]
    \centering
   \includegraphics[width=0.9\textwidth]{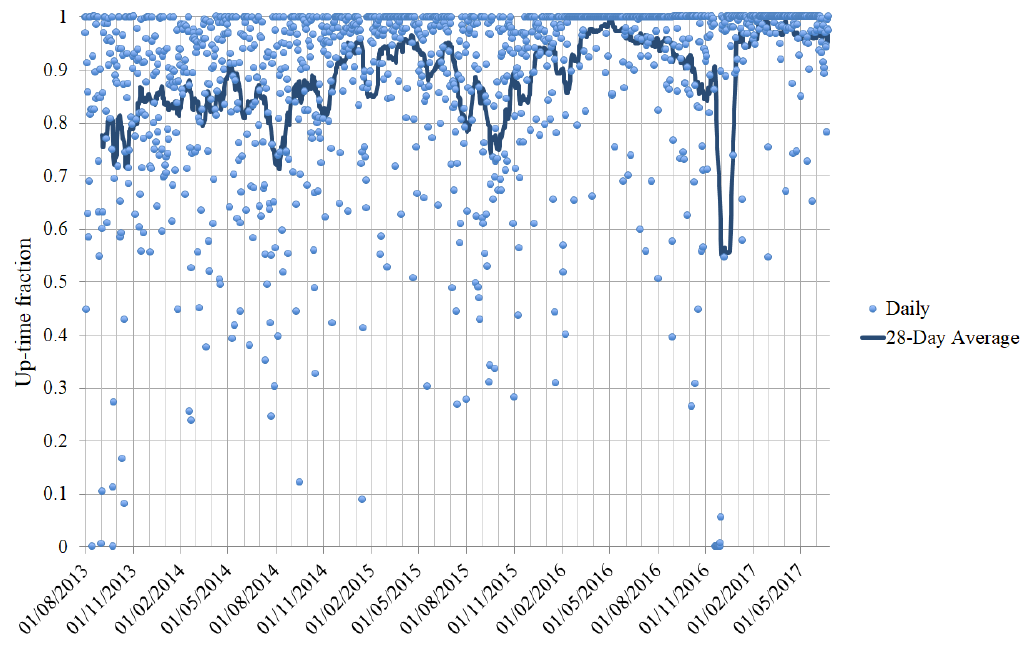}
    \caption{Uptime fraction as a function of time. Uptime has steadily increased over time as we move from commissioning to steady state running.}
   \label{fig:uptime}
  \end{figure}

For every particle physics experiment, the success depends on the monitoring of the detector performance. There are various tools developed to track the detector performances. The high quality of data is monitored by performance metrics developed through automated script. One metric that is used to gauge the overall performance of the detectors is their fractional uptime, shown in Fig.\ref{fig:uptime}. Maximizing uptime is a top priority for the experiment, because while one may have any number of sophisticated tools for reconstructing neutrino kinematics, they are of little use if the data was never recorded in the first place.

\section{Identifying Issues With Live Data : Control Room Tools}
\begin{figure}[!htbp]
    \centering
   \includegraphics[width=0.9\textwidth]{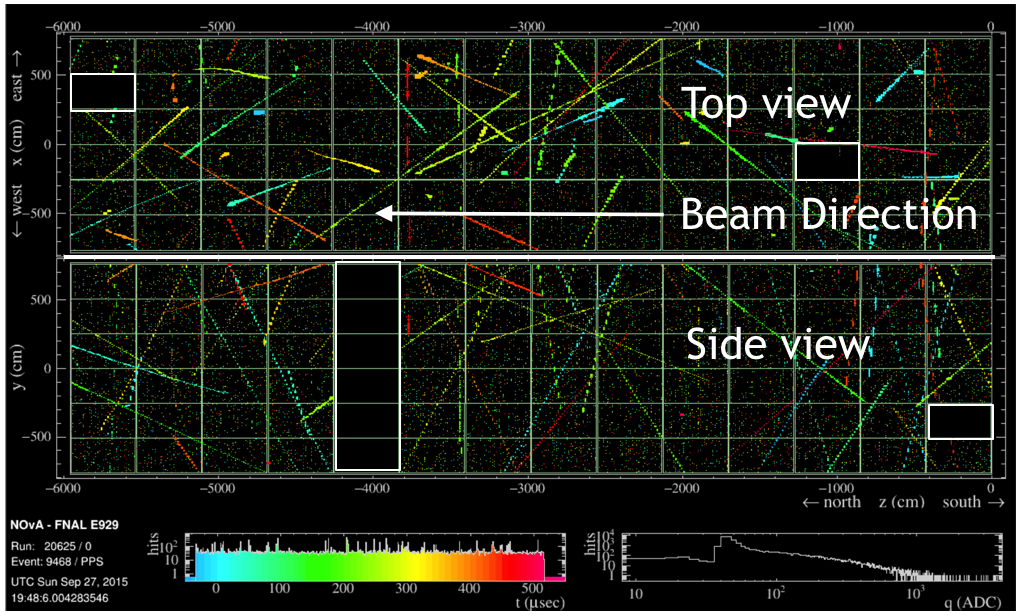}
    \caption{An event shown in the NOvA FD which collects all hits recorded in a 550 microsecond window and some of the non-reporting DCMs (white box).}

   \label{fig:evd}
  \end{figure}

The raw hits collected through various electronic systems and propagated through the data acquisition system (DAQ) is displayed in a ``human-readable'' way in the control room and this is monitored by the shifter. A live event display \cite{evd} provides a closer look at the real-time state of data-triggered events for the purpose of a high-level overview of detector performance. There are several tools we have designed to identify faulty behavior part of our detector. One of the control room tools is an event display which easily catch non-reporting DCMs or channels (Fig.\ref{fig:evd}) as well as detector timing problems can create broken tracks, which are identified easily by looking at a few cosmic ray tracks that span several DCMs. 
  
\begin{figure}[!htbp]
    \centering
   \includegraphics[width=0.85\textwidth]{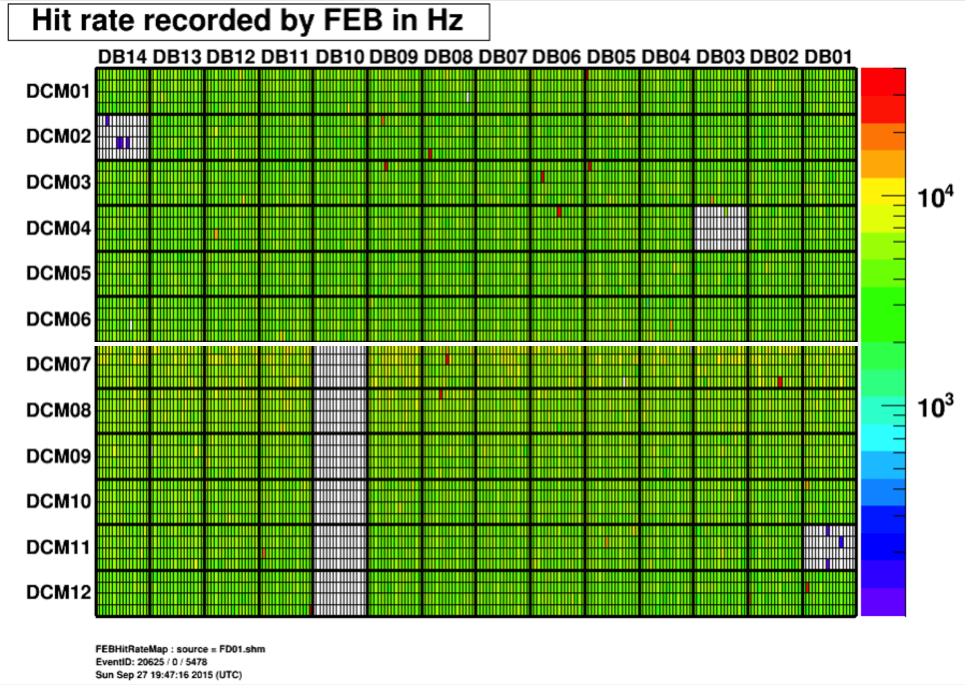}
    \caption{As the hit information comes in from raw data, hit rate recorded by FEB in Hz and some of the non-reporting DCMs (empty).}

   \label{fig:onmon}
  \end{figure}

The raw event contains all of the information for the whole event. This includes the time and total charge for that hit and the hardware address in terms of FEB and pixel number. The nanoslices are associated with individual channels while microslices are associated with individual DCMs. Each hit is categorized by the charge (ADC): ``low'' (ADC $<$ 175), ``MIP'' (175 $\leqslant$ ADC $\leqslant$ 3200), and ``high'' (3200 $<$ ADC). A hit rate plot (shown in Fig.\ref{fig:onmon}) is computed by dividing the hit counts by the total live time; this is another diagnosis tools. The hit count histograms are filled continuously and the hit rates are recomputed after every 100 processed events. 
  
\begin{figure}[!htbp]
    \centering
   \includegraphics[width=0.85\textwidth]{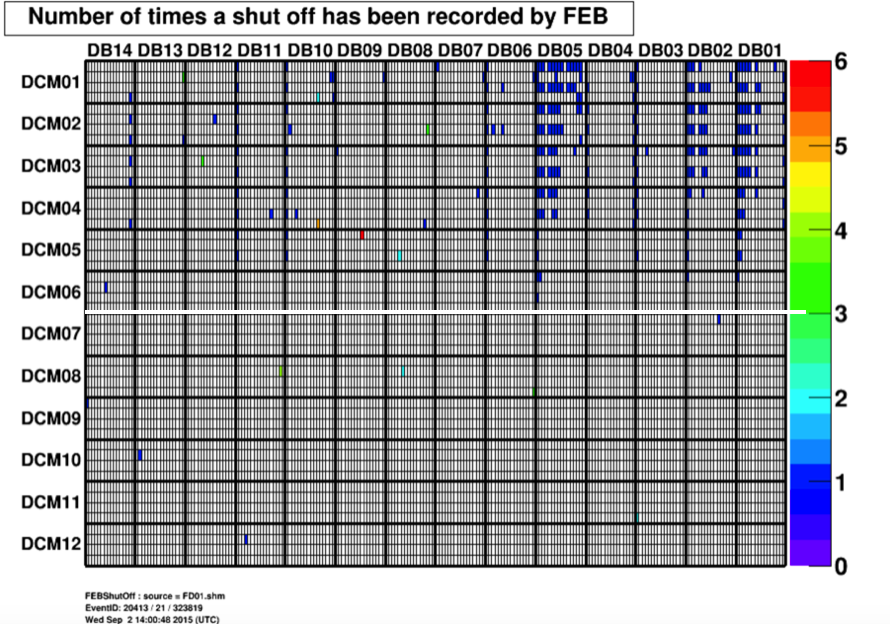}
    \caption{Number of times a shut off has been recorded by FEB.}
   
   \label{fig:dropout}
  \end{figure}

Another metric that keeps track of how many times each FEB switches from a reporting state to a non-reporting is called ``FEB dropout''. The number of dropouts is recorded and these FEBs are added to list for possible replacement. Sometimes if there is thunderstorm, this can triggered multiple drop outs and we can easily catch the non-reporting part of the detector as shown in Fig.\ref{fig:dropout}.

\section{Data Quality Metrics}
\begin{figure}[!htbp]
    \centering
   \includegraphics[width=0.9\textwidth]{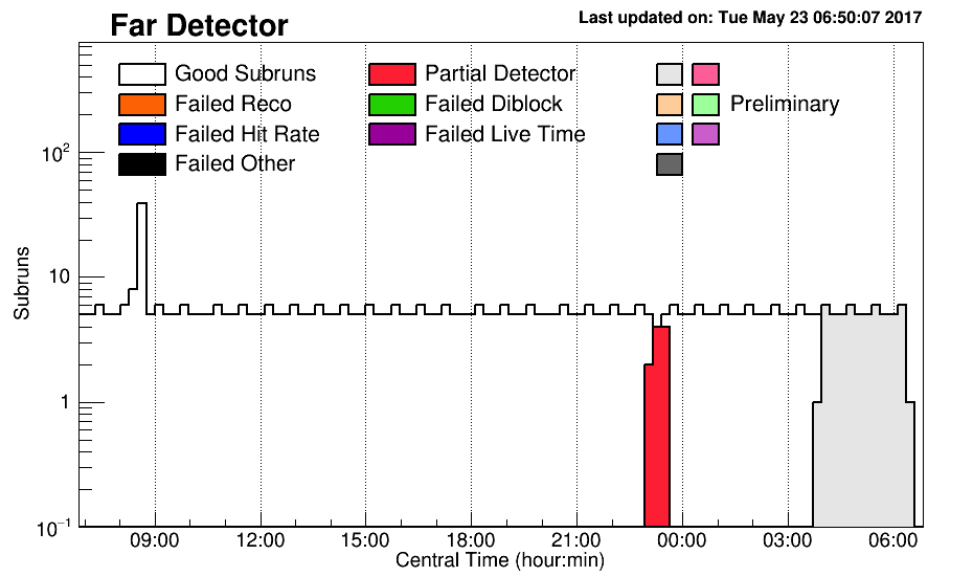}
    \caption{Good subruns as a function time, where different colors show different data quality failures. Good (white), too many or too few tracks (orange), median hit rate too high or low (blue), part of detector missing or has too high or low hit rate (red and green), no activity or incorrect time stamps (black), fewer than 1k triggers in run or not enough data to access quality (violet).}
   \label{fig:dataquality}
  \end{figure}

\begin{figure}[!htbp]
    \centering
   \includegraphics[width=0.9\textwidth]{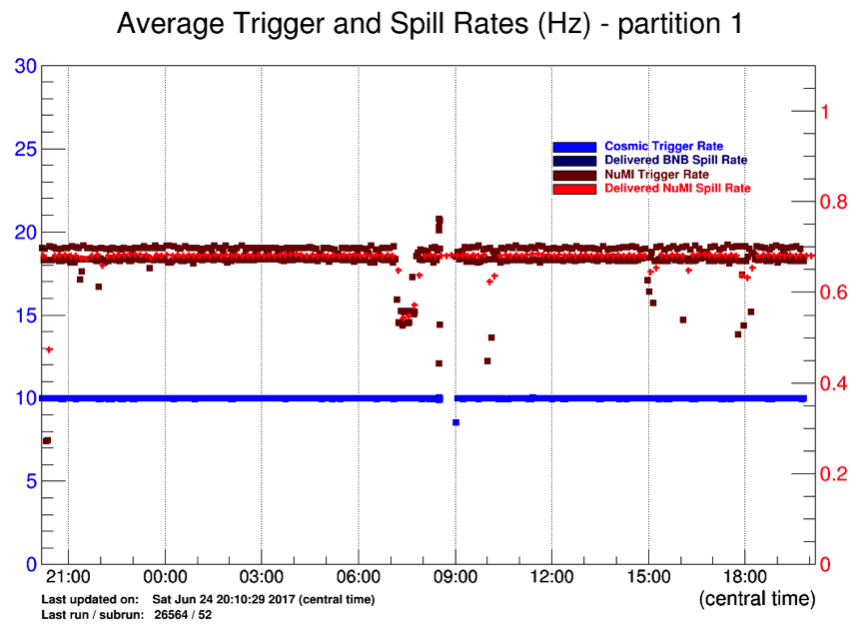}
    \caption{A combined plot of average NuMI and cosmic trigger rates per subrun. Also included is the beam spill rate (every 1.33s we receive 10 microsecond of beam) which should directly correspond to NuMI trigger rate. If the difference between the beam spill rate and recorded NuMI trigger rate is grater than 0.05 Hz or if the NuMI trigger rate exceeds 0.8 Hz the plot turns red which is easily spotted by the shifter.}
   
   \label{fig:spill}
  \end{figure}

There is a series of fully automated scripts that handles everything from data processing to transferring data to a permanent disk. We run multiple data quality \cite{nearline} checks over the data to generate performance plots that are regularly published to a website spanning time intervals of hours, days, weeks, months, and years. For the permanent record, high data quality decisions are made including a determination of which subruns (a collection of events over a period of time) represent good data and which channels should be masked prior to reconstruction. There are different data quality failures as shown in Fig.\ref{fig:dataquality}. Another metric to track cosmic and NuMI beam spill trigger rate is shown in Fig.\ref{fig:spill}. 

\section{Automating Maintenance List due to Hardware issues}
\begin{figure}[!htbp]
    \centering
   \includegraphics[width=0.45\textwidth]{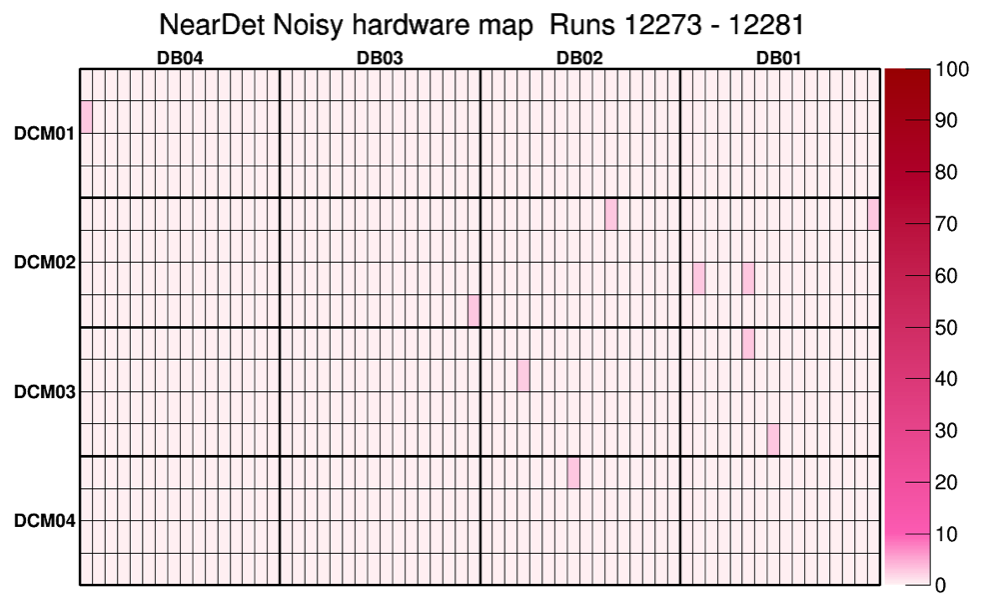}
    \includegraphics[width=0.45\textwidth]{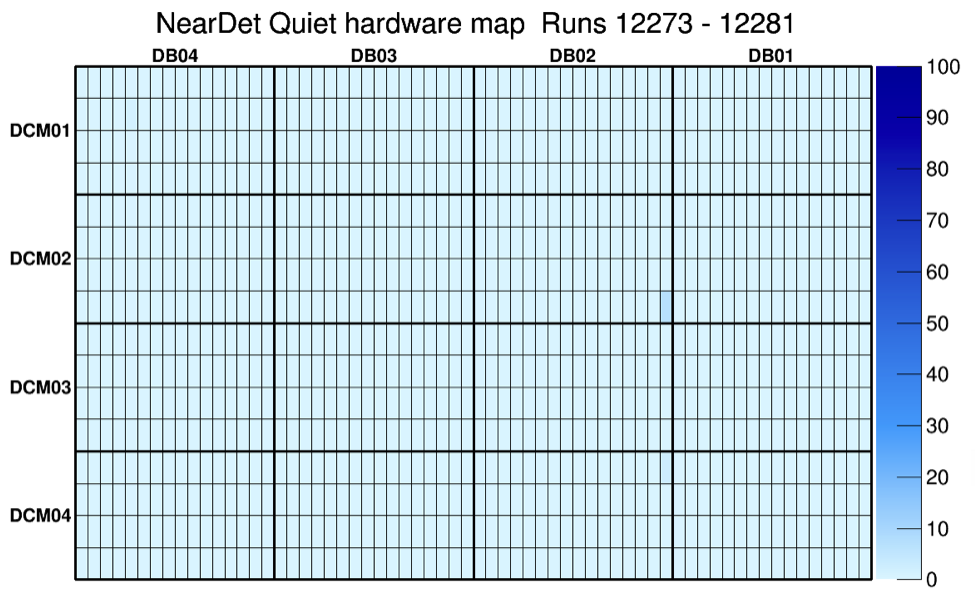}
    \caption{The percentage of time each channel has reported a higher (left)  and lower (right) than normal rate.}
   \label{fig:maintanance}
  \end{figure}

\begin{figure}[!htbp]
    \centering
   \includegraphics[width=0.9\textwidth]{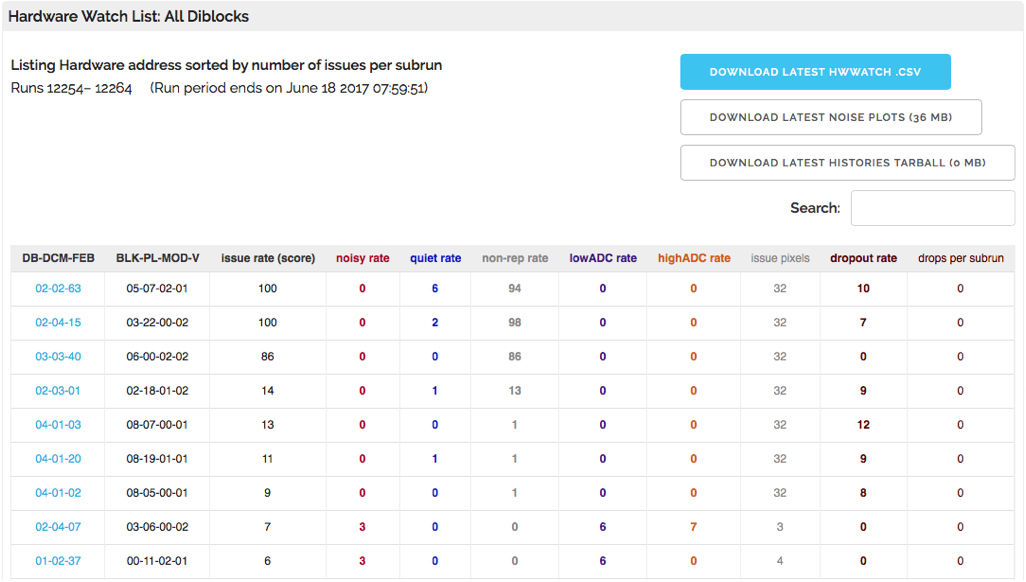}
    \caption{The watch list keeps track of various signs that hardware failures. Once the issue rate becomes high enough the component is placed.}
   \label{fig:list}
  \end{figure}

The hit rate and drop out metrics produced by the online monitoring are useful indicators of issues in our readout electronics. A watch list of hardware with potential issues is computed weekly based on these metrics. An example of a noisy hardware map and a hardware map after maintenance is shown in Fig.\ref{fig:maintanance}. Also a script generates the list of ``issue channels'' which is the main driver of scheduled maintenance tasks on the APDs and FEBs in both the Near and Far detectors. An example of this list of issued channel are shown in Fig.\ref{fig:list}.

\section{Summary}
Since October 2013, the far detector has been taking data and the near detector, since August 2014. We continuously monitor the data recorded, both in real time in the control room and after recording to remove any data with potential data quality issues. Tracking detector performance and data quality is a key aspect of NOvA data-taking as it helps to maintain a high detector uptime fraction and dictates the data selection for all physics analyses. This note highlighted several tools developed and utilized for diagnosis of detector failures. Uptime has steadily increased over time as we move from commissioning to steady state running and is now consistently above 95\%. 

\section{ACKNOWLEDGMENTS}
This work was supported by the US Department of Energy; the US National Science Foundation; the Department of Science and Technology, India; the European Research Council; the MSMT CR, Czech Republic; the RAS, RMES, and RFBR, Russia; CNPq and FAPEG, Brazil; and the State and University of Minnesota. We are grateful for the contributions of the staffs of the University of Minnesota module assembly facility and NOvA FD Laboratory, Argonne National Laboratory, and Fermilab. Fermilab is operated by Fermi Research Alliance, LLC under
Contract No. DE-AC02-07CH11359 with the U.S. Department of Energy, Office of Science, Office of High Energy Physics. 








\begin{thebibliography}{60}


\bibitem{tdr}
D.S. Ayres et al. The nova technical design report. Technical report, Fermilab, Batavia, Illinois, 2007.

\bibitem{nearline}
https://nusoft.fnal.gov/nova/datacheck/nearline/nearline.html

\bibitem{evd}
http://nusoft.fnal.gov/nova/public/

\end{thebibliography}
\end{document}